\documentclass[conference,Letter]{IEEEtran}
\usepackage{graphicx,times,cite,amsmath, amssymb, epsfig, array}
\usepackage{array}
\usepackage{textcomp}
\usepackage[noend]{algorithmic}             
\usepackage{algorithm}               
\usepackage{multirow}                
\usepackage{amsmath}
\usepackage{xcolor}
\usepackage{booktabs}

\newtheorem{definition}{Definition}

\newlength{\figwidth}
\newcommand{\tabincell}[2]{\begin{tabular}{@{}#1@{}}#2\end{tabular}}
\begin{document}
\title{On the necessity of separating MAC protocols into data and control planes}

\author{\IEEEauthorblockA{Zhongjiang~Yan*, Bo~Li, Mao~Yang}\\
\IEEEauthorblockA{School of Electronics and Information, Northwestern Polytechnical University, Xi'an 710072, China\\ Email: \{zhjyan, libo.npu, yangmao\}@nwpu.edu.cn}}

\maketitle
\begin{abstract}
The network protocol architecture not only can be designed from the traditional view of layers, but also can be designed from the view of planes, i.e., the data, control and management planes. Media access control (MAC) is a function of the data link layer, and thus the MAC protocols involve of both the data and control planes. However, although the international wireless local area network (WLAN) standard, IEEE 802.11 or Wi-Fi, has developed over 20 years, the control plane of the MAC protocols is not explicitly described yet. Thus, does it need to separate  the MAC protocols into data and control planes? If not, are there some problems in existing hybrid architecture? To answer above questions, we analyse the possible problems of the current MAC protocols in IEEE 802.11, particularly in std 802.11-2020. These problems can be seen as new starts for the next study of the WLAN for the next generation (WNG).
\end{abstract}

\section{Introduction}
IEEE 802.11\cite{Std80211by2020} is an international standard designed for the wireless local area network (WLAN), which mainly focus on the media access control (MAC) and physical (PHY) layers, which are two functions specified in the open system interconnect (OSI) reference model. From the layered view of OSI model, PHY of IEEE 802.11 is responsible to carry a bit stream from the source to the destination over a wireless link, and to deal with the mechanical and electrical specifications of the antennas and the wireless transmission media, i.e., the wireless propagation channel. On the other part, as a main part of the data link layer in the OSI model, MAC is responsible to deal with the function of how to share the wireless media among the neighbours. And thus a lot of rules, i.e., the MAC protocols, are designed in IEEE 802.11 to regulate the behaviours of the nodes when accessing into the channels. Above layered viewpoint is well known by the industrial and academic area researchers.

However, besides the layered viewpoint, the OSI model also can be viewed from the planes\cite{Matthew80211acSurvivalGuide2013}, i.e., the data, control and management planes.  The management plane provides protocols to allow the network administrator to configure and monitor the network elements. As for the data and control planes, there exists a coupled relationship between them since both of them are involved in the MAC protocols. On one hand, the protocols in the data plane are responsible to move bits from one location to another, or to move frames from the input interfaces to the output interfaces. On the other hand, by changing the behaviours of the protocols of the data plane according to the dynamic environment, the control plane helps the network to operate smoothly. For example, in WLAN, the control plane can help mobility between access points (APs), coordinating channel selection, authenticating users, coordinating the media access behaviours of the nodes and so on. 

The related works on the development of the data and control planes are given as follows, mainly focusing on the evolution development of WLAN in IEEE 802.11. Fig. 1.8 of Ref. \cite{Perahia80211nac2013} summarizes the MAC enhancements of 802.11n and 802.11ac, where these enhancements are given from the data, control and management planes, respectively. It can be found that not only the throughput and robustness enhancing features, but also a number of optional extensions cross both the data and control planes. These numerous optional features in 802.11n and 802.11ac mean that extensive signalling of device capability is required to ensure coexistence and interoperability. For example, AP needs to manage the wide channel BSS so that different channel width devices are able to associate with the BSS and operate.  Furthermore, some extensions cross both the control and management planes, for example, the phased coexistence operation (PCO) and low power supports, i.e., power save multi-poll (PSMP).

Although the frame formats of the MAC are classified into three types, i.e., the data, control and management frames, in the published IEEE 802.11-2020 standard\cite{Std80211by2020}, but only the data plane is clearly described. Fig. 5.1 of  std 802.11-2020\cite{Std80211by2020} shows the MAC data plane architecture, while the control plane architecture are not included. The data plane involves of the processes of transporting of all or part of a MAC service data unit (MSDU) and a MAC management protocol data unit (MMPDU). In other words, the management plane is also coupled with the data plane where the MMPDUs are sent within the data plane. However, the handshakes of the control frames, for example, request to send (RTS) and clear to send (CTS), neither belong to MSDU nor MMPDU, which should belong to the control plane. In other words, the processes of the control frames transmission and reception are not involved in Fig. 5.1. As a consequence, the MAC control plane are not explicitly presented and described.

However, the data and control frames are deeply coupled in the MAC protocols. For example, in the mandatory distributed coordination function (DCF) of 802.11, if the size of the MPDU waiting to be transmitted is larger than a given threshold, say 500 Bytes, then the node should employ the two-way handshake at first, i.e., RTS-CTS handshake, before sending the data. Otherwise, the node can send the data directly after the backoff process finishes successfully. Furthermore, as the development of the standard, the control plane of 802.11 becomes further complicated due to the new features' introduction, for example, multi-user MAC (MU-MAC) and restricted target wakeup time (rTWT). Along with them, the traditional contention based 802.11 MAC evolves as a hybrid one, which combines the contention based, scheduling based and the reservation based method together. Therefore, it is time to think about whether it is necessary to separate the MAC protocols into the data and control planes respectively, before moving to study of the WLAN for the next generation (WNG).

In this paper, we focus ourselves on the relationships of the MAC protocols' data and control planes, and systematically analyse the necessaries to separate MAC protocols into the data and control planes, respectively. Particularly, 6 classical problems are listed and taken as examples to illustrate the problems of the current hybrid mode of the MAC protocols. These problems can be taken as new start points of WNG, and as the targets of the next generation MAC protocols.

The followings of this paper are organized as follows. Section \ref{s:defandrelations} formally defines the MAC protocols' data and control planes, and discusses their relationships in current std 802.11-2020\cite{Std80211by2020}. Section \ref{s:problems} details 6 problems of current MAC protocols of std 802.11-2020\cite{Std80211by2020}, and the possible reasons of these problems are summaries and discussed. Section \ref{s:disscusonsolution} discusses possible solutions to separate the MAC control plane from the data plane. Section \ref{s:conclustion} concludes this paper.

\section{Definition and relationships}\label{s:defandrelations}
\subsection{Definitions}
\begin{definition}
\textit{\textbf{MAC protocols' data plane (DP).}} DP is defined as the data frames' transmission and reception processes of the MAC protocols, where the data frame is packaged in the form of MAC's data or management frame format
\end{definition}

Note that the definition of DP in this paper follows with the description of the MAC data plane architecture of Std 802.11-2020 \cite{Std80211by2020} in Fig 5.1. In other words, DP is responsible for transmission and reception of high layer's user data and management data, by employing obtained reliable transmission opportunity (TXOP) by CP. Classical reliable data transmission MAC protocols include Stop and Wait automatic repeat request (ARQ), Go-Back-N ARQ and Selective Repeat ARQ, the details of which are referred to Ref. \cite{ForouzanDataComm2012}. Based on above traditional mechanisms, the MAC protocols regulated in IEEE 802.11 develop with the introduction of the new technologies. For example, to efficiently utilizing the obtained reliable TXOP, the MAC protocols employ technologies in terms of reduced inter-frame space (RIFS) burst, aggregated MPDU (AMPDU) in 802.11n, and multi-user resource unit (RU) in orthogonal frequency division multiple access (OFDMA) in 802.11ax. 

\begin{definition}
\textit{\textbf{MAC protocols' control plane (CP).}} CP is defined as the control frames' transmission and reception processes of the MAC protocols, where the control frame is packaged in the form of MAC's control frame format. 
\end{definition}

Note that the definition of CP is similar with that of DP, defined in this paper, where the only difference is that the type of frame format is control in CP. In other words, CP is responsible for establishing reliable data transmission links, or for obtaining reliable transmission opportunity (TXOP) for the DP over the shared wireless transmission media, by employing the capabilities offered by the PHY layer. For example, to reduce the collision effects on the long data, the handshake of RTS-CTS is employed. To increase the efficiency of high data rate link, block acknowledgement (BA) mechanism is introduced, where the BA handshake should accomplished before the AMPDU transmission and reception. To reduce the collision effects of the high density user scenario, the trigger (TRG) mechanism and the reservation based mechanism, i.e., TWT, are introduced.

\textit{Remarks.} From above definitions, one can find that there exist clear boundaries of DP and CP, respectively. However, we would like to note that in the processes flows of the MAC protocols the control frame handshake processes always inter-leave with the data transmission and reception processes. For example, in the four-way handshake of RTS-CTS-DATA-ACK, only DATA is in the data frame format while the other frames are in the control frame format. As another example, in the transmission of AMPDU, the BA mechanism should be always employed explicitly or implicitly, immediately or deferred. Therefore, there exists deeply coupled relationships between DP and CP. Thus, before moving to exampling the problems of the current inseparate DP and CP, their relationships are classified and detailed as follows.

\subsection{Relationships}
The coupled relationships of the current inseparate DP and CP can be illustrated as two dependence relationships, i.e., the timing and casual dependence relationships between CP and DP. They are detailed  in sequence.
\subsubsection{Timing dependence relationships}
Taking the simplest DCF as an example, since CP and DP work on the same channel, thus it can be seen that there exists two timing dependence relationships between CP and DP, which include two cases.
\begin{itemize}
\item \textbf{Immediately DP after CP.} In DCF, there exists a packet length threshold. If the packet length is larger than the threshold, then it should employ the four way handshake, i.e., RTS-CTS-DATA-ACK. The handshake of RTS-CTS belongs to CP, and that of DATA-ACK belongs to DP. Furthermore, it can be seen that in DCF and EDCA whenever the CP succeeds the DP follows immediately.
\item \textbf{Directly DP without CP.} Following with the packet length threshold philosophy, if the packet length is smaller than the threshold, the small packets can be send directly, i.e., employing the two-way handshake DATA-ACK, or directly DP without CP. 
\end{itemize}

Therefore, from the bird view the channel shows an interleaving timing relationship dependence between CP and DP. 
\subsubsection{Causal dependence relationship}
Due to different objectives, CP and DP show different characters, i.e., reliable CP and high rate DP. 
\begin{itemize}
\item \textbf{Reliable CP.} Usually, CP employs low MCS, a few of spatial streams and maximum transmission power, to reliably obtain the shared transmission opportunity (TXOP), such that the data and management frame can be reliably transmitted. 
\item \textbf{High rate DP.} Different from CP, DP usually employs as high as possible MCS and as many as possible number of MIMO spatial streams, such that the data rate of the wireless link will be as high as possible.
\end{itemize}

The primary reasons resulting in the differences are as follows. CP aims to broadcast the control frames to every neighbourhood nodes inside the BSS, such that it can reliably obtain the shared TXOP and avoid hidden terminal problems. Thus, the TXOP can be uniquely utilized by the pair of sender and receiver. However, DP aims to unicast the data from the source to the destination, by high efficiently exploiting the obtained TXOP.  That is because DP holds the belief that the obtained TXOP is uniquely belonging to the sender and the receiver themselves. Therefore, there exists a causal dependence relationship between CP and DP. In other words, the high rate DP is depended on the reliable CP. If the CP can not be reliably guaranteed, the reliability of DP will not be guaranteed, not to mention the high data rate of DP.

\section{Problems of existing inseparated IEEE 802.11 MAC protocols}\label{s:problems}
Table \ref{t:summary} summaries the analysed problems and possible reasons which result in them. Totally speaking, there are 6 problems proposed in this paper, which are resulted from the inseparative DP and CP\footnote{Note that there are several viewpoints to analyse the problems, such as following the development of the standard, i.e., the sequence of new introduced features, }.  Each of the problems is listed and exampling in sequence.
\begin{table*}
\centering
\caption{Summary of problems and reasons}\label{t:summary}
\begin{tabular*}{0.85\linewidth}{@{}cll@{}}
\toprule
No.	&  Problems	&  Reasons\\
\midrule 
1 	& Collisions of intra CPs and between CP and DP	 &  \tabincell{l}{•	Hidden terminals\\ •	loss of NAV} \\ \hline
2  & Blockings of intra DPs and between CP and DP&\multirow{2}{*}{•	Low rate CP and long time DP exist on the same channel or link.}\\ \cline{1-2}
3	& Bottleneck of primary channel&\\ \hline
4	& Low usage of secondary channels&  \tabincell{l}{•	Bottleneck of primary channel\\
•	Channel binding capabilities/rules\\
•	Independent secondary channels' states\\
} \\ \hline
5	& Relatively reduced efficiency of CP 	&  \tabincell{l}{•	Increased bandwidth of DP\\
•	Increased PHY rate of DP}\\ \hline
6	& Reduced efficiency of DCF and EDCA	&  \tabincell{l}{•	Fragmented access intervals caused by new schemes, e.g., rTWT}\\
\bottomrule 
\end{tabular*}
\end{table*}

\subsection{Collisions of intra CPs and between CP and DP}
\begin{figure}
\centering
\includegraphics[scale=0.6]{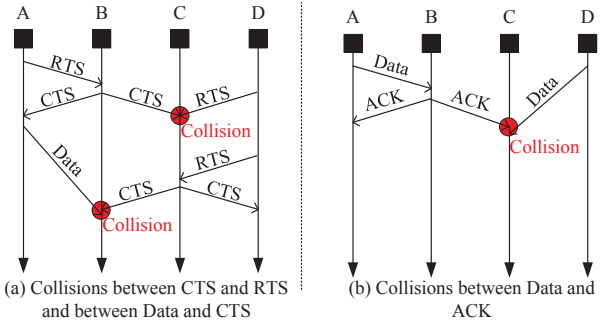}
\caption{Example of CP and DP collision problem.}
\label{f:dcfCollision}
\end{figure}

Fig. \ref{f:dcfCollision} shows an example of collisions of intra CPs and between CP and DP. It can be seen that there are two kinds of collisions occurred in the basic DCF. 
\begin{enumerate}
\item \textbf{Intra-collision of CP.} The first one is the collision between node $B$'s CTS and node $D$'s RTS, as shown in Fig.\ref{f:dcfCollision} (a), which can be seen as CP's intra-collision.
\item \textbf{Inter-collision between CP and DP.} The second collision is the one between node $C$'s CTS and node $A$'s Data, as shown in Fig.\ref{f:dcfCollision} (a),  which can be seen as the inter-collision between CP and DP. As another example, in Fig.\ref{f:dcfCollision} (b), even though RTS-CTS handshake is not employed, node $B$'s ACK collides with node $D$'s Data. 
\end{enumerate}

As shown in Table \ref{t:summary}, the possible reasons of the collision problem include two parts, i.e., hidden terminals and loss of network allocation vector (NAV). The first one is the hidden terminal problem. As shown in Fig.\ref{f:dcfCollision} (a), the collision between node $B$'s CTS and node $D$'s RTS occurs because node $D$ does not receive the RTS from node $A$ between which the distance is 3 hops. Then, node $D$ sends an RTS which collides with the CTS sending from node $B$ at node $C$. The second one is due to loss of NAV. As shown in Fig.\ref{f:dcfCollision} (a), the collision between node $C$'s CTS and node $A$'s Data occurs, because node $C$ does not keep quiet during the data transmission between nodes $A$ and $B$ due to previous collision and missing the NAV contained in node $B$'s CTS. As another example, as shown in Fig.\ref{f:dcfCollision} (b), the collision between $B$'s ACK $D$'s Data occurs because node $D$ and $C$ are not aware of the reception process of node $B$, since no NAV information is broadcast in advance.

\textit{Remarks.} Note that the examples in Fig.\ref{f:dcfCollision} only follows with the rules of DCF, which is the mandatory mechanism of IEEE 802.11. Lots of extended mechanisms are based on DCF, for example, the enhanced distributed coordination access (EDCA). However, since DCF is a MAC designed for distributed network and it mainly works for CP, thus if the DP is not separated from DCF the reliability and high rate of DP can not be guaranteed. In other words, CP can continue to employ the DCF-based mechanisms, while to improve the reliability and the data rate of the DP, other mechanisms can be helpful, for example, the scheduling based and the reservations based mechanisms. This can be seen the first possible way to separate the MAC protocols into CP and DP.

\subsection{Blockings of intra DPs and between CP and DP}
\begin{figure}
\centering
\includegraphics[scale=0.6]{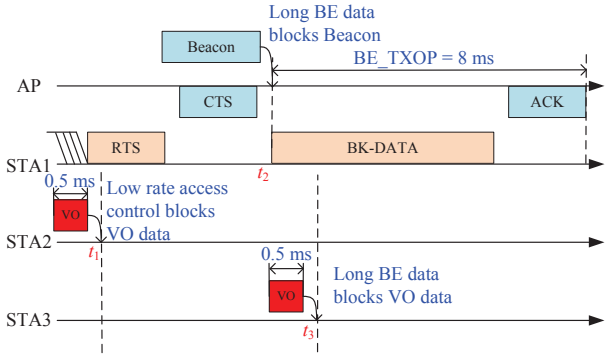}
\caption{Example of CP and DP blocking problem.}
\label{f:dcfBlock}
\end{figure}

Fig. \ref{f:dcfBlock} shows an example of blockings of intra DPs and between CP and DP. There are 3 cases of blocking occurred, detailed as follows.
\begin{enumerate}
\item \textbf{Low rate CP blocks low latency DP.} At $t_1$, a low latency voice (VO) data arrives at STA2, which is a small data, e.g., with transmission time length $0.2ms$. However, it finds the channel is busy since STA1 is transmitting RTS, and thus it must defer the transmission of VO data. Since the RTS frame is transmitting in a low rate, thus it can be seen as a blocking caused by low data CP to low latency DP.
\item \textbf{Long DP blocks the management frame.} At $t_2$, a Beacon frame is generated at AP, which should be periodically broadcast. However, STA1 begins its background (BK) data transmission, where the obtained transmission opportunity (TXOP) is a long time duration, e.g., $8ms$, and thus AP must defer the broadcasting of the Beacon frame. Since the data transmission duration is much longer, thus it can be seen as a blocking caused by long BE data to the management frame, i.e., the Beacon frame.
\item \textbf{Inter-blocking in DP.} At $t_3$, a VO data arrives at STA3. However, it finds the channel is busy since STA1 is transmitting a long BE data, and thus it must defer the transmission of VO data. Therefore, it can be seen as a blocking caused by long BE data to the VO data, i.e., inter blocking in DP.
\end{enumerate}

As shown in Table \ref{t:summary}, the possible reason of the blocking problem is that	low rate CP and long time DP exist on the same channel or link. This is also the possible reason of the bottleneck problem of the primary channel, which will be detailed therein.
\subsection{Bottleneck of primary channel}
\begin{figure}
\centering
\includegraphics[scale=0.45]{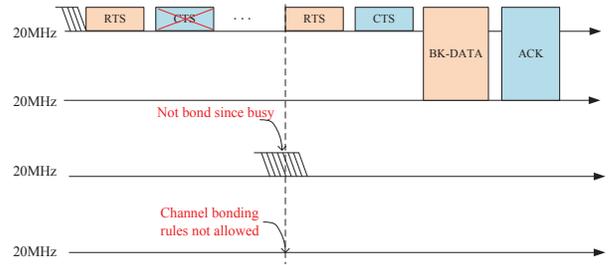}
\caption{Example of channel binding problems.}
\label{f:channelBinding}
\end{figure}

Fig. \ref{f:channelBinding} shows an example of bottleneck of primary channel, when channel binding mechanism is employed. In the channel binding mechanism, as regulated in Std 802.11-2020 \cite{Std80211by2020}, one channel is specified as the primary channel, where the bandwidth may be 20MHz, 40MHz or 80MHz, according to the channel binding mode. And the other channels are specified as the secondary channels. 

Whenever the sender wants to access into the channel and transmit data, it should handshake with the receiver first through the primary channel. If the handshake successes, then the data can be transmitted through the primary channel, and optionally through the secondary channel if which is idle. In other words, the primary channel not only is used by CP for multiple access handshake, but also is used by DP for data transmission. Therefore, if there are many users, such as in the density scenario, the primary channel will be very busy due to employing by not only CP but also DP, which will become the bottleneck of the whole system such that will reduce the efficiency of the whole system.

\textit{Remarks.} As shown in Table \ref{t:summary}, the basic reason of the blocking problem as shown in Fig. \ref{f:dcfBlock} and the bottleneck problem as shown in Fig. \ref{f:channelBinding} is that low rate CP and long time DP exist on the same channel or link. Furthermore, since the MAC protocols regulated in Std 802.11-2020 \cite{Std80211by2020} are asynchronous mechanisms, and there are no boundaries between the CP and DP, i.e., not separated in the current standard, thus the problems of blocking and contention between them appear. Therefore, to avoid or to solve above two problems, some kind of boundaries between them should be defined and regulated. For example, some time durations or channels are specified to CP or DP exclusively. This can be seen the second possible way to separate the MAC protocols into CP and DP.

\subsection{Low usage of secondary channels}
Continuing the discussion on Fig. \ref{f:channelBinding}, whenever the bottleneck problem of the primary channel occurs, no matter how many secondary channels there exist they can not be employed. What's more, the more secondary channels the lower usage of them.

As shown in Table \ref{t:summary}, besides of the bottleneck problem of the primary channel, there are two possible remaining reasons, i.e., the channel binding capabilities and the independent secondary channels' state. For the first one, assume that the AP is capable of channel binding 160MHz while the STA is only capable of 40MHz, then most of the secondary channels on the AP side can not be utilized, resulting in low usage of the secondary channels. For the second one, assume that both AP and STA are capable of 80MHz. However, when the channel is binding together the third one becomes busy, due to independent channel state. This results in the third and the four 20MHz channel can not be binded together with the first two ones, due to the current channel binding rules. In other words, the independent secondary channels' state can also be seen as another appearance of the blocking problem where another transmission link blocks this one.

\textit{Remarks.} As shown in Table \ref{t:summary}, there are 3 different reasons which result in the problem of low usage of secondary channels. As aforementioned discussion, these reasons can also be analysed from other viewpoints. However, if there is a tag which can be assigned to the MAC protocols, one can assign the channel binding mechanism as the DP but not the CP. This can be seen the third possible way to separate the MAC protocols into CP and DP.

\subsection{Relatively reduced efficiency of CP }
\begin{figure}
\centering
\includegraphics[scale=0.6]{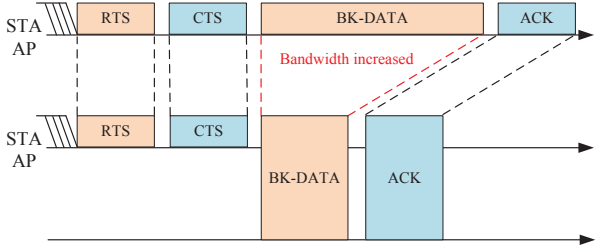}
\caption{Example of increased bandwidth reducing CP efficiency.}
\label{f:increasedBandwidth}
\end{figure}

Fig. \ref{f:increasedBandwidth} continues the discussion on channel binding mechanism, which shows an example of increased bandwidth reducing CP efficiency. The behind casual dependence is that whenever the channel binding mechanism succeeds the bandwidth of the DP is enlarged, but that of CP remains. Larger bandwidth higher data rate, and shorter time duration. And thus, the time duration ratio of DP over CP becomes reduced. In other words. the efficiency of CP is relatively reduced since the data rate of DP is increased due to larger bandwidth.

\begin{figure}
\centering
\includegraphics[scale=0.6]{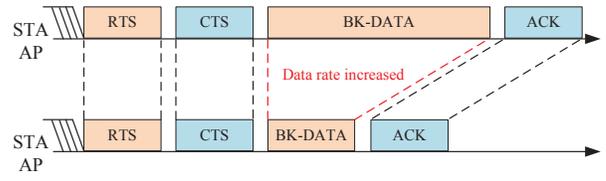}
\caption{Example of increased data rate reducing CP efficiency.}
\label{f:increasedDataRate}
\end{figure}

Fig. \ref{f:increasedDataRate} shows another example of increased data rate reducing CP efficiency. From the development of the standard, one can see that the data rate of PHY continues increasing. For example, the modulation scheme is 64QAM in 802.11n, which is increased to 256QAM in 802.11ac, to 1024QAM in 802.11ax and to 4096QAM in 802.11be. Therefore, the data rate of PHY continues increasing from 802.11n to 802.11be. However, this increasing only contributes to DP, but not the CP, since that CP usually employs the basic MCS to ensure the reliability of data transmission. Higher data rate shorter data transmission duration. Similarly, the time duration ratio of DP over CP becomes reduced.

\textit{Remarks.}  As shown in Table \ref{t:summary}, the basic reasons which relatively reduce the efficiency of CP include two parts, i.e., the increased bandwidth and PHY rate of CP. However, if one's viewpoint moves further one can seen that the basic reason can be concluded as the timing dependence of the CP and DP. In other words, both in Fig. \ref{f:increasedBandwidth} and Fig. \ref{f:increasedDataRate}, the CP and DP have a deeply coupled timing dependence over the same channel, i.e., immediately DP after CP. Therefore, the radio of the time duration of each determines the efficiency of them. To overcome this problem, the timing dependence between CP and DP should be decoupled such that CP and DP can be separated. This can be seen the fourth possible way to separate the MAC protocols into CP and DP.

\subsection{Reduced efficiency of DCF and EDCA}
\begin{figure}
\centering
\includegraphics[scale=0.45]{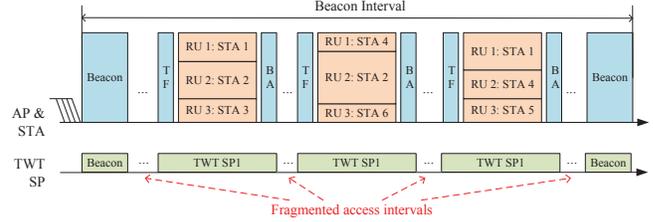}
\caption{Example of fragmented access intervals.}
\label{f:fragmented}
\end{figure}

Fig. \ref{f:fragmented} shows an example of fragmented access intervals of rTWT, which reduces the efficiency of DCF and EDCA. As a new introduced feature, the rTWT mechanism reserves many service periods (SPs) over the time line for many different transmission links. These SPs divide the time line into different parts and DCF-based mechanisms can only access into the channel during the fragmented access intervals of rTWT. However, due to the bounded duration of the intervals, the case may occur that only a little time is left after successful channel access, and thus this reduces the efficiency of the DCF-based mechanisms.

\textit{Remarks.}  As shown in Table \ref{t:summary}, the basic reason to reduce the efficiency of the DCF-based mechanisms is the fragmented access intervals. However, if one's viewpoint moves further one can seen that the basic reason can be concluded as the casual dependence of the CP and DP in the DCF-based mechanisms. In other words, if the DP is immediately after the CP, as that in DCF-based mechanisms, the bounded time duration will constrict the time length of the DP. To overcome this problem, the casual dependence of the CP and DP should be decoupled, such that CP and DP can be separated. This can be seen the fifth possible way to separate the MAC protocols into CP and DP.

\section{Discussions on possible solutions}\label{s:disscusonsolution}
Following the analysed problems of existing inseparated IEEE 802.11 MAC protocols, the main reasons and possible solutions can be summarised as follows.
\begin{enumerate}
\item To avoid the collision problems, CP can continue to employ the DCF-based mechanism due to the contention based mechanism of unlicensed frequency, while the DP can employ the schedule based or the reservation based mechanism to improve the reliability and the data rate of the data transmission.
\item  To avoid the blocking and bottleneck problems, some kind of boundaries between the CP and DP can be defined and regulated, for example, the specified time durations or channels to CP and/or DP.
\item To avoid the low usage of the secondary channels, some kind of tags can be defined and are assigned to the new and existing features, to regulate how to use these features in CP and DP, respectively.
\item To avoid the efficiency of CP relatively reduced, the timing dependence between CP and DP can be decoupled, i.e., avoiding the immediately DP after CP.
\item To avoid the efficiency of DCF-based mechanisms reduced, the casual dependence between CP and DP can be decoupled, i.e., avoiding incorporating DP into the DCF-based mechanism.
\end{enumerate}
In summary, it is necessary to separate the MAC protocols into CP and DP to improve the efficiency of the MAC protocols, and to improve the quality of service. For the CP, the DCF-based mechanism can continue to be employed, but avoiding to incorporate the DP into it to break the timing and casual dependences between CP and DP. For the DP, the scheduling based and the reservation based mechanisms can be employed, such that the boundaries of the CP and DP can be defined and regulated. 

Note that in the open literatures from the academic researchers, there are lots of contribution in the area of scheduling and reservation based MAC protocols. For example, Ref. \cite{LiBo2010mStepReservation} studies the performance of the multiple step reservation base MAC protocols, which shows that the reliability of the reservation can be largely improved even using only 2 or 3 steps of reservation. Ref. \cite{yangbo2019singleradio} proposes a multi-channel based reservation MAC protocol for single radio, which shows that the reservation information can be guaranteed to be received even for the single radio with sophisticated design of the MAC process flows. Other useful clues can be found therein, which is out of the scope of this paper.

\section{Conclusion}\label{s:conclustion}
The MAC protocols not only can be viewed as a basic sub-layer of the data link layer in OSI model, but also can be viewed as a combination of the data, control and management of planes. However, the international WLAN standard, Std 802.11-2020, does not separate the MAC protocols into data and control planes yet. Existing six classical problems are analysed based on the current MAC protocols' architecture, and the possible reasons which result in the problems are analysed. Five possible methods can be employed to overcome the proposed problems, where the basic ideas are to break the timing and casual dependence between CP and DP. For the separated CP and DP, it is suggested the CP continues to employ the DCF-based mechanisms but avoiding incorporate DP into it, and the DP can be implemented with the scheduling based and the reservation based mechanisms. Useful clues can be found in Refs. \cite{LiBo2010mStepReservation} and \cite{yangbo2019singleradio}, and the references therein.

\section*{Acknowledgments}
This work was supported in part by the National Natural Science Foundations of CHINA (Grant No. 61771392, 61771390 and 61871322).

\end{document}